\documentstyle[pra,preprint,aps]{revtex}
\begin{document}
\def\veps{\varepsilon}
\newcommand{\eref}[1]{(\ref{#1})}
\tighten \draft

\title{$E$1 amplitudes, lifetimes, and polarizabilities
of the low-lying levels of atomic ytterbium}

\author{S.~G.~Porsev, Yu.~G.~Rakhlina, and M.~G.~Kozlov}
\address{Petersburg Nuclear Physics Institute, Gatchina, Leningrad
         district, 188350, Russia}
\date{\today}
\maketitle

\begin{abstract}
The results of {\it ab initio} calculation of $E$1 amplitudes, lifetimes,
and polarizabilities for several low-lying levels of ytterbium are
reported. The effective Hamiltonian for the valence electrons $H_{\rm
eff}$ has been constructed in the frame of CI+MBPT method and solutions
of many electron equation $H_{\rm eff} \Phi_n = E_n \Phi_n$ are found.

\end{abstract}

\pacs{PACS. 31.15.Ar, 32.10.Dk, 32.10.Fn}

\narrowtext

\section*{I. Introduction}

In this paper we report results of an {\it ab initio} calculation of
$E$1 amplitudes, lifetimes, and polarizabilities for several low-lying
levels of ytterbium. In Ref.~\cite{PRK} we calculated the energies and
hyperfine structure (hfs) constants of low-lying levels of ytterbium. In
that calculations the accuracy of atomic wave functions was tested at
the short distances by comparison of the calculated hfs constants with
the experimental ones. The latter are usually known to a very good
accuracy, providing a good test of the quality of the wave function near
the nucleus.

$E$1 amplitudes, in contrast, are determined by the wave function behavior
at large distances. Usual experimental accuracy for the oscillator
strengths and scalar polarizabilities is on the level of few percent.
This is close or even less than the accuracy of precise atomic
calculations (see, e.~g., calculations for Ba \cite{KP98} and Cs
\cite{DFS97}). Tensor polarizabilities can be measured with the accuracy
of 1\% or better \cite{KO,JAP}. Thus, it is possible to test an atomic
wave function at large distances at 1\% level. Note that 1\% accuracy is
crucial for calculations of parity nonconservation effects in atoms,
because it allows to test predictions of the Standard model at small
momentum transfer \cite{DFS97,MR}. So far such precision has been
achieved only for one-electron atoms Cs and Fr \cite{DFS89,BJS,DFS95}.
In this work we deal with a much more complicated Yb atom.

We consider ytterbium as a two electron atom with the core
[$1s^2,...,4f^{14}$]. Valence-valence correlations are taken into
account by the configuration interaction (CI) method, while core-valence
and core-core correlations are treated within the second order of the
many-body perturbation theory (MBPT). The latter is used to construct an
effective Hamiltonian for the CI problem in the valence space. The
details of the method can be found in the papers \cite{DFK,KP}.
Application of this method to calculations of hfs constants has been
discussed in \cite{PRK,KP98,DFKP}. In Ref.~\cite{KP98} the method has
been extended to the calculations of polarizabilities. Here we apply
this technique for calculations of lifetimes, $E$1 amplitudes, and
polarizabilities of ytterbium.

\section*{II. General formalism}
 Let us write out several formulae that will be used in the following.
 The expression for oscillator strength for
 $a, J \rightarrow a^\prime, {J^\prime}$ transition has the form
 \cite{Sobel} (atomic units $m=\hbar=e=1$ are used throughout the
 paper):
\begin{equation}
     f(a J, a^\prime {J^\prime}) =
   - \frac{2~\omega_{a J, a^\prime {J^\prime}}}{3~(2J+1)}
     |\langle a, J||D||a^\prime, {J^\prime} \rangle|^2,
\label{a1}
\end{equation}
 where $\omega_{a J, a^\prime {J^\prime}} = E_{a J} - E_{a^\prime
 {J^\prime}}$, $\bf D$ is the dipole moment operator,
 and reduced matrix elements (MEs) are defined as follows:
\begin{eqnarray}
        &&\langle a',J',M'|D_q|a,J,M\rangle
\nonumber \\
        &&= (-1)^{J'-M'}
        \left( \begin{array}{ccc} J' & 1 & J \\ -M' & q & M
        \end{array} \right)
        \langle a',J'||D||a,J\rangle.
\label{a2}
\end{eqnarray}
  The lifetime $\tau$ of a level is the inverse of the total transition
rate. The probability for $a, J \rightarrow a^\prime, J^\prime$
transition is given by:
\begin{eqnarray}
  W(a J, a^\prime {J^\prime}) = \frac{4}{3 c^3}
    \frac{\omega_{a J, a^\prime J^\prime}^3}{2J+1}
     |\langle a, J||D||a^\prime, J^\prime \rangle|^2,
\label{a3}
\end{eqnarray}
where $c$ is the speed of light.

Static polarizability of the sublevel $|a,J,M\rangle$ in a DC electric
field ${\bf E} = {\cal E} \hat{\bf z}$ is defined as:
\begin{eqnarray}
        \Delta E_{a,J,M} &=& - \frac{1}{2} \alpha_{a,J,M} {\cal E}^2
\\
        &=& - \frac{1}{2} \left(
        \alpha_{0,a,J} + \alpha_{2,a,J}
        \frac{3M^2-J(J+1)}{J(2J-1)} \right) {\cal E}^2,
\label{a4} \nonumber
\end{eqnarray}
where $\Delta E_{a,J,M}$ is the energy shift and $\alpha_0$ and
$\alpha_2$ define the scalar and tensor polarizabilities,
correspondingly. Being a second order property, $\alpha_{a,J,M}$ can be
expressed as a sum over unperturbed intermediate states:
\begin{equation}
        \alpha_{a,J,M} = -2 \sum_{n}
        \frac{|\langle a,J,M|D_z|n,J_n,M\rangle|^2}{E_{a}-E_{n}},
\label{a5}
\end{equation}
where $E_n$ is an unperturbed energy of a level $n$, and the sum runs
over all states of opposite parity. The formalism of the reduced MEs
allows to write explicit expressions for the scalar and tensor parts of
the polarizability:
\begin{eqnarray}
        \alpha_{0,a,J} &=& \frac{-2}{3(2J+1)}
        \sum_{n} \frac{|\langle a,J||D||n,J_n\rangle|^2} {E_{a}-E_{n}},
\label{a6} \\
        \alpha_{2,a,J} &=&
        \left(\frac{40J(2J-1)}{3(2J+3)(2J+1)(J+1)}\right)^{1/2}
\label{a7} \\
        &&\times \sum_{n} (-1)^{J+J_n+1}
        \left\{ \begin{array}{ccc} J & 1 & J_n \\ 1 & J & 2
        \end{array} \right\}
        \frac{|\langle a,J||D||n,J_n\rangle|^2} {E_{a}-E_{n}}.
\nonumber
\end{eqnarray}
In order to use Eqs.~\eref{a5}--\eref{a7} in calculations one needs to
know a complete set of eigenstates of the unperturbed Hamiltonian. It
becomes practically impossible when dimension of a CI space exceeds few
thousand determinants. It is known, that it is much more convenient to
solve inhomogeneous equation instead of the direct summation over the
intermediate states \cite{Stern,Dalg}. Indeed, let us consider the
solution of the following equation:
\begin{eqnarray}
        &&(E_a - H)|X_{a,M'} \rangle = D_q |a,J,M \rangle,
\label{a8}
\end{eqnarray}
where $q=0,\pm 1$ and $M'=M+q$. Obviously, the right hand side in
Eq.~\eref{a5} can be expressed in terms of the function $X_{a,M}$ (note
that $D_0 \equiv D_z$):
\begin{eqnarray}
        \alpha_{a,J,M} &=& -2 \langle a,J,M |D_0| X_{a,M} \rangle.
\label{a9}
\end{eqnarray}

If we want to rewrite Eqs.~\eref{a6} and \eref{a7} in terms of the
function $X_{a,M'}$, we need to decompose the latter in terms that
correspond to particular angular momenta $J_i$. Generally speaking,
there can be three such terms with $J_i=J,J \pm 1$:

\begin{eqnarray}
        && X_{a,M'} = X_{a,J-1,M'} + X_{a,J,M'} + X_{a,J+1,M'}.
\label{a10}
\end{eqnarray}

Now, with the help of the functions $X_{a,J',M'}$ Eqs.~\eref{a6} and
\eref{a7} are reduced to:
\begin{eqnarray}
        \alpha_{0,a,J} &=& (-1)^{q+1} \frac{2}{3(2J+1)}
\label{a11}\\
        &&\times \sum_{J'}
        \left( \begin{array}{ccc} J' & 1 & J \\ -M' & q & M
       \end{array} \right)^{-2}
        \langle a,J,M|D_{-q}|X_{a,J',M'}\rangle,
\nonumber \\
        \alpha_{2,a,J} &=& (-1)^{q+1}
        \left(\frac{40J(2J-1)}{3(2J+3)(2J+1)(J+1)}\right)^{1/2}
\nonumber \\
        && \times \sum_{J'} (-1)^{J+J'}
        \left\{
        \begin{array}{ccc} J & 1 & J' \\ 1 & J & 2 \end{array}
        \right\}^{-2}
\label{a12}\\
        &&\times \left( \begin{array}{ccc} J' & 1 & J \\ -M' & q & M
        \end{array} \right)^{-2}
        \langle a,J,M|D_{-q}|X_{a,J',M'}\rangle,
\nonumber
\end{eqnarray}
where sums run over $J'=J,J \pm 1$. Note, that these equations are valid
only if all $3j$-symbols on the right hand side do not turn to zero. One
has to take it into account when choosing for what spherical component
$q$ to solve Eq.~\eref{a8}.

If we know the solution of Eq.~\eref{a8} and its decomposition
\eref{a10}, then expressions \eref{a11} and \eref{a12} allow us to find
both scalar and tensor polarizabilities of the state $|a,J \rangle$.
Moreover, the same functions $X_{a,J',M'}$ can be also used to find
other second order atomic properties, such as amplitudes of the
Stark-induced $E$1 transitions or parity nonconserving $E$1 transitions
between the states of the same nominal parity (see, for example,
Ref.~\cite{KPF}).

\section*{III. Calculation details and results}
\subsection{Orbital basis set and CI space.}

The calculation procedure is quite similar to that of Ref.~\cite{KP98}.
For this reason we give here only a brief description of its features.
This calculation is done in the $V^N$ approximation, that means that
core orbitals are obtained from the Dirac-Hartree-Fock (DHF) equations
for a neutral atom (we use the DHF computer code \cite{BDT}). The basis
set for the valence electrons includes 6s, 6p, 5d, 7s, 7p, 6d DHF
orbitals and 8s--15s, 8p--15p, 7d--14d, 5f--10f, and 5g--7g virtual
orbitals.  The latter were formed in two steps. On the first step we
construct orbitals with the help of a recurrent procedure, which is
similar to that suggested in Ref.~\cite{Bogdan} and described in
Refs.~\cite{KP,KPF}. After that we diagonalize the $V^N$ DHF operator to
obtain the final set of orbitals.

For this orbital basis set the complete CI is made for both even-parity
and odd-parity levels. Two-electron wave functions are the linear
combinations of the Slater determinants with a given $J_z$. It means
that no symmetrization with respect to angular momentum $J$ is made.

\subsection{Effective operators.}
Within the CI+MBPT method the wave function of the valence electrons is
found from the eigenvalue equation:
\begin{eqnarray}
        && H_{\rm eff}|a,J,M \rangle
        = E_a |a,J,M \rangle.
\label{b00}
\end{eqnarray}
Eq.~\eref{a8} is rewritten as equation for valence electrons
only:
\begin{eqnarray}
        &&(E_a - H_{\rm eff})|X_{a,M'} \rangle
        = D_{{\rm eff},q} |a,J,M \rangle,
\label{b0}
\end{eqnarray}
with the effective operators, which are found by means of the MBPT. The
effective Hamiltonian for two valence electrons is formed within the
second order MBPT \cite{DFK}. We used RPA for the effective dipole
moment operator (see, for example, Ref.~\cite{MP}). We have checked that
MBPT corrections to ${\bf D}_{\rm eff}$, which are not included in RPA,
are small if RPA equations are solved with 6s electrons excluded from
the self-consistency procedure. That means that RPA equations have the
same form as in the $V^{N-2}$ approximation. The more detailed
description of the effective operator formalism is given in \cite{DFKP}.

\subsection{Transition amplitudes and lifetimes.}
We first solve eigenvalue
Eq.~\eref{b00} with the effective Hamiltonian for low-lying even-
and odd-parity states.  Strictly speaking, the effective Hamiltonian can
be safely used only for the energy levels below the core excitation
threshold. For Yb this threshold lies at 23189~cm$^{-1}$ above the
ground state \cite{Martin}. However, it was shown in \cite{PRK} that 
theoretical spectrum is quite good up to $\sim$40000~cm$^{-1}$. 
Correspondingly, we can work (with some caution) with the states lying 
slightly above the core excitation threshold. In our approach we fail 
to reproduce the states with unfilled $f$ shell and correspondingly to 
account properly for the interaction with such states. For this reason 
we restrict ourselves to the consideration of the states lying 
sufficiently far from those with unfilled $f$ shell. We consider $E$1 
transitions between four low-lying odd-parity states 
($^3P_{0,1,2}^o\,(6s6p)$ and $^1P_1^o\,(6s6p)$) and seven even-parity 
states ($^1S_0\,(6s^2)$, $^3D_{1,2,3}\,(5d6s)$, $^1D_2\,(5d6s)$, 
$^3S_1\,(6s7s)$, and $^1S_0\,(6s7s)$). The state $^1P_1^o\,(6s6p)$ 
requires special attention. The nearest $f^{13}\,5d\,6s^2$ state lies 
only 3800 cm$^{-1}$ above the latter and their interaction is not 
negligible. We estimated that configuration $f^{13}\,5d\,6s^2$ 
contributes on the level of several percent to the wave function of 
$^1P_1^o\,(6s6p)$ state. We do not take into account this configuration 
mixture. This reduces the accuracy of the calculated $^1P_1^o\,(6s6p) 
\rightarrow {}^1L_J$ $E$1 amplitudes.

When eigenfunctions for the valence electrons are found, we can
calculate transition amplitudes and lifetimes. The results of
calculations are presented in Table~\ref{tab1}. The magnitudes of the
$E$1 amplitudes vary in a wide range. These variations correspond in 
part to the approximate selection rules $\Delta S=0$ and $\Delta J = 
\Delta L$, which are easily traced through Table~\ref{tab1}. For large 
amplitudes we estimate the accuracy of our calculation to be 3-5\%. For
the reason discussed above the amplitudes
$\langle{}^1L_J|D|^1P_1^o\,(6s6p) \rangle$ do not follow this rule. The
accuracy for these amplitudes, as well as for small amplitudes ($\leq
0.5$~a.u.), is about 15-20\%.

Where available, we compare our results with those of other theoretical
\cite{Magda,Migd} and experimental
\cite{Budk,BW,Penkin,Andersen,Cris,Bai} groups. For the convenience of
comparison we recalculated the oscillator strengths and transition
probabilities to the reduced MEs. Calculations in Ref.~\cite{Magda} were
performed in the $L$-$S$ coupling scheme. The simplest semiempirical method
\cite{Bates} was used then to evaluate the radial parts. In
Ref.~\cite{Migd} the multiconfiguration Dirac-Fock method was used. The
valence-core electronic correlations were included semiempirically.
Comparing our results with the results of other theoretical works, one
can see that it was important to account for the valence-core
correlations.

Now, using Eq.~\eref{a3} we can find the transition probabilities and
the lifetimes of the levels (see Table~\ref{tab2}). In these
calculations we used experimental transition frequencies. Therefore, the
accuracy of these numbers depends only on the accuracy of the dominant
transition amplitudes. As a result, the largest error (40\%) takes place
for the states ${}^1S_0$(6s7s) and ${}^1D_2$(5d6s) where the transition
to the state ${}^1P^o_1(6s6p)$ is dominant.  For other states we
estimate theoretical accuracy for the lifetimes as 10\% or better.

\subsection{Polarizabilities.}
In order to find the polarizabilities we substitute eigenfunctions in
the right hand side of Eq.~\eref{b0} and solve corresponding
inhomogeneous equation. After that Eqs.~\eref{a11} and \eref{a12} give
us $\alpha_0$ and $\alpha_2$. Results of these calculations are
presented in Table~\ref{tab3}. It is seen that, unlike barium (see
Ref.~\cite{KP98}), $\alpha_2$ has typically the same order of magnitude
as $\alpha_0$. For this reason the theoretical accuracy for $\alpha_2$,
as a rule, is similar to that for $\alpha_0$. In contrast, experimental
data for $\alpha_2$ are usually much more precise and complete.

There are several sources of errors in the calculations of
polarizabilities. Some of them are the same as for hfs calculations, and
are connected with the inaccuracy in the wave functions and the
effective operators (note, that RPA corrections to the dipole operator
are much smaller than for hfs operators). The additional source of
errors is the inaccuracy in eigenvalues. Finally, solving Eq.~\eref{b0}
we do not account for configurations
$4f^{13}\,nl\,n^{\prime}l^{\prime}\,n^{\prime \prime}l^{\prime \prime}$.
In other words, we do not take into account $f$ shell polarization.
Fortunately, the states of configuration
$4f^{13}\,nl\,n^{\prime}l^{\prime}\,n^{\prime \prime}l^{\prime \prime}$,
that can be reached by one-electron transitions from the levels studied
here, lie rather far above the latter. The estimates show, that
contribution of $f$ shell polarization to the polarizabilities of the
states listed in Table~\ref{tab3} does not exceed 2-3 a.u.

The final accuracy of calculations is very different for different
levels. For instance, the 95\% of the polarizability of the ground state
$^1S_0\,(6s^2)$ is due to the ME $\langle ^1S_0|D|{}^1P_1^o\,(6s6p)
\rangle$. Supposing that this ME is calculated within the accuracy of
20\%, the latter for $\alpha_0\,({}^1S_0)$ will be about 40\%
(corresponding transition frequency is reproduced almost ideally
\cite{PRK}). We need to say that even taking into account the large
uncertainty of our result, it significantly differs from
$\alpha_0\,(^1S_0)\,=\,266$ a.u. obtained in Ref.~\cite{KMS} where the
Hartree-Fock method was used.

For the $D_J\,(5d6s)$ states the situation is more complicated. There
are large cancellations between contributions of $P_J^o\,(6s6p)$ states
and higher-lying states. For this reason their polarizabilities are
small and the role of different small contributions is enhanced. Thus,
analysis of the accuracy becomes difficult; only for the tensor
polarizability of $^3D_3\,(5d6s)$ state we can estimate the accuracy
to be 20\%. All other values of $\alpha_0$ and $\alpha_2$ for
$D_J\,(5d6s)$ states presented in Table~\ref{tab3} are rather estimates
by the order of magnitude.

The scalar polarizabilities of the levels $^3S_1\,(6s7s)$ and
$^1S_0\,(6s7s)$ are basically determined by the MEs $\langle
S_{J^\prime}\,(6s7s)|D|P_J^o\,(6s7p) \rangle$. Because of the closeness
of $f^{13}\,5d^2\,6s$ states we failed to obtain the reliable wave
functions for $P_J^o\,(6s7p)$ states. Correspondingly, the values for
$\alpha_0(^3S_1\,(6s7s))$ and $\alpha_0(^1S_0\,(6s7s))$ are also only
the estimates.

Now let us go over to the odd-parity states. The accuracy of $\alpha_0$ 
and $\alpha_2$ for $^3P_J^o\,(6s6p)$-triplet is 6-10\%. The main 
contribution here comes from the $^3D_J\,(5d6s)$ multiplet and there
are no cancellations because all important levels of opposite parity lie
above and contribute with the same sign. The accuracy for $\alpha_0$ of
$^1P_1^o(6s6p)$ state is about 40\% and for $\alpha_2$ even worse
(50\%). This is due to the large contribution of the intermediate state
$^1S_0\,(6s7s)$ to these polarizabilities (see above).

  In Ref.~\cite{Li} the Stark shift of the $^1S_0\,(6s^2) \rightarrow
{}^3P_1^o\,(6s6p)$ transition in ytterbium was measured. The Stark shift
rate was found to be K= $-$61.924\,(0.193) a.u.  In terms of
polarizabilities this magnitude can be written as:
$$K= -\frac{1}{2} \left\{ \alpha_0\,(^3P_1^o) -
2\alpha_2\,(^3P_1^o)-\alpha_0\,(^1S_0) \right\}.$$
Using the numbers from Table~\ref{tab3}, we find that K= $-$55\,(9)
a.u., in good agreement with the experimental result \cite{Li}.

In Ref.~\cite{Budk}, the Stark shifts for $^1S_0\,(6s^2) \rightarrow
{}^3D_{1,2}\,(5d6s)$ transitions were observed. These shifts depend on
the differences in scalar polarizabilities $\left(\alpha_0\,({}^1S_0) -
\alpha_0\,({}^3D_{1,2})\right)$:
\begin{eqnarray}
\alpha_0\,(^1S_0) - \alpha_0\,(^3D_1) &=& \left\{
\begin{array}{ll}  71  & \mbox{theory,}\\
                   86\,(3) & \mbox{experiment,}
\end{array} \right. \nonumber  \\
\alpha_0\,(^1S_0) - \alpha_0\,(^3D_2) &=& \left\{
\begin{array}{ll}  82  & \mbox{theory,}\\
                   80\,(4) & \mbox{experiment,}
\end{array} \right. \nonumber
\end{eqnarray}
where theoretical values are taken from Table~\ref{tab3}.

The method used here allows us to calculate not only static
polarizabilities, but, also the Stark-induced amplitudes for different
transitions. For instance, the magnitude of the vector transition
polarizability $|\beta|$ for $^1S_0\,(6s^2) \rightarrow
{}^3D_1\,(5d6s)$ transition was calculated to be 122\,(12) a.u., in good
agreement with our previous calculation 138\,(30) a.u. \cite{Porsev} and
experimental result 113\,(15) a.u. \cite{Budk}.

\subsection{Conclusion.}
Application of the effective operator technique to the Yb atom is
hampered by the existence of the shallow 4$f$ shell. Nevertheless, it is
possible to make reliable calculations of different atomic properties
including transition frequencies, hyperfine constants, $E$1 amplitudes,
lifetimes and polarizabilities for many low-lying energy levels. It is
of a particular importance, that with some caution calculations can be
done even for levels above the core excitation threshold, which for Yb
lies at 23189~cm$^{-1}$.

\section{Acknowledgments}

This work was supported in part by Russian Foundation for Basic
Research, Grant No.~98-02-17663. One of us (SP) is grateful to
the St.~Petersburg government for financial support,
Grant No.~M98-2.4P-522.

\newpage
\mediumtext
\begin{table}
\caption{Reduced MEs
     $|\langle L_J||r||L^\prime_{J^\prime} \rangle|$ (a.u.).
     Calculations were made in the $L$-gauge. Other theoretical
     and experimental results are given where available. The
     uncertainties are indicated in the parentheses.}

\label{tab1}

\begin{tabular}{cllll}
& \multicolumn{1}{c}{$^3P_0^o(6s6p)$} &
\multicolumn{1}{c}{$^3P_1^o(6s6p)$} &
\multicolumn{1}{c}{$^3P_2^o(6s6p)$} &
\multicolumn{1}{c}{$^1P_1^o(6s6p)$} \\
\hline
$^1S_0(6s^2)$ &---& 0.54\,(8)  & ---        & 4.40\,(80)   \\
              &   & 0.44\tablenotemark[1]  && 4.44 \tablenotemark[1]\\
              &   & 0.549\,(4)\tablenotemark[3]  && 4.89
                                                    \tablenotemark[2]\\
              &   & 0.553\,(13)\tablenotemark[4] && 4.13\,(10)
                                                    \tablenotemark[4]\\
              &   &                              && 4.02
                                                    \tablenotemark[5]\\
              &   &                              && 4.26
                                                    \tablenotemark[6]\\
$^3D_1(5d6s)$ & 2.61\,(10)  & 2.26\,(10)  &0.60\,(12)  & 0.27\,(10)  \\
&             &  2.2\,(1) \tablenotemark[7]   &&0.24 \tablenotemark[1]\\
$^3D_2(5d6s)$ & ---   & 4.03\,(16)  &2.39\,(10)  & 0.32\,(6)  \\
&       &                                    && 0.60\tablenotemark[1] \\
$^3D_3(5d6s)$ & ---   & ---   & 6.12\,(30)  & ---   \\
$^1D_2(5d6s)$ & ---   & 0.54\,(10)  & 0.38\,(8)  & 3.60\,(70)  \\
$^3S_1(6s7s)$ & 1.98\,(10)  & 3.53\,(15)  & 5.05\,(20)  & 0.73\,(15)\\
& 1.36 \tablenotemark[2] & 2.50 \tablenotemark[2]
                              & 3.77 \tablenotemark[2] & \\
$^1S_0(6s7s)$ & ---   & 0.22\,(4)  & ---   & 4.31\,(80)  \\
&             &         0.22\,(2) \tablenotemark[8] && \\
\end{tabular}
Theory: \tablenotemark[1]{Ref.~\cite{Migd},}
\tablenotemark[2]{Ref.~\cite{Magda};}

Experiment: \tablenotemark[3]{Ref.~\cite{Budk},}
\tablenotemark[4]{Ref.~\cite{BW},}
\tablenotemark[5]{Ref.~\cite{Penkin},}
\tablenotemark[6]{Ref.~\cite{Andersen},}
\tablenotemark[7]{Ref.~\cite{Cris},} \tablenotemark[8]{Ref.~\cite{Bai}.}
\end{table}
\narrowtext
\begin{table}

\caption{Lifetimes (nsec) of the low-lying levels for Yb calculated with
the reduced MEs from Table~\ref{tab1} and experimental transition
frequencies. }

\label{tab2}

\begin{tabular}{llcc}
State  & Config.     & This work     & Other data      \\ \hline
$^3D_1$ &  $5d6s$  &   372\,(30)   &   380\,(30) \tablenotemark[1]  \\
$^3D_2$ &  $5d6s$  &   430\,(35)   &   460\,(30) \tablenotemark[1]  \\
$^3D_3$ &  $5d6s$  &   540\,(55)   &                                \\
$^1D_2$ &  $5d6s$  &  4400\,(1800) & 6700\,(500) \tablenotemark[1]  \\
$^3S_1$ &  $6s7s$  &  13.5\,(1.1)  & 12.5\,(1.5) \tablenotemark[2]  \\
          &          &               & 15.9\,(1.9) \tablenotemark[3]  \\
$^1S_0$ &  $6s7s$  &  33\,(13)     & 45.8\,(1.0) \tablenotemark[4]  \\
\hline $^3P_1^o$ & $6s6p$ &  875\,(250) &  760--875 \tablenotemark[5]
\\
            &        &             &  1294     \tablenotemark[6]  \\
$^3P_2^o$ & $6s6p$ & 15.0\,(1.5) sec & 14.5 sec \tablenotemark[6]   \\
$^1P_1^o$ & $6s6p$ &  5\,(2) &  5.1--6.4  \tablenotemark[5] \\ &
&             & 4.78                    \tablenotemark[6] \\
\end{tabular}
Experiment: \tablenotemark[1]{Ref.~\cite{Cris},}
\tablenotemark[2]{Ref.~\cite{WLange},}
\tablenotemark[3]{Ref.~\cite{Baum},} \tablenotemark[4]{Ref.~\cite{Bai},}
\tablenotemark[5]{see Ref.~\cite{Blagoev} and references therein;}

Theory: \tablenotemark[6]{Ref.~\cite{Migd}.}

\end{table}
\mediumtext
\begin{table}

\caption{Scalar and tensor polarizabilities (a.u.) of low-lying levels
of Yb. Theoretical accuracy is indicated where analysis was possible,
otherwise the numbers should be considered as estimates.}

\label{tab3}

\begin{tabular}{llccc}
&&\multicolumn{2}{c}{Theory}
 &\multicolumn{1}{c}{Experiment} \\
level  & config.     &\multicolumn{1}{c}{$\alpha_0$}
                     &\multicolumn{1}{c}{$\alpha_2$}
                     &\multicolumn{1}{c}{$\alpha_2$} \\
\hline $^1S_0$ &  $6s^2$  &    118\,(45)  &
                     &                          \\
$^3D_1$ &  $5d6s$  &    47        &     22
                                    &     28\,(4) \tablenotemark[1]  \\
$^3D_2$ &  $5d6s$  &    36        &     17
                                    &     28\,(8) \tablenotemark[1]  \\
$^3D_3$ &  $5d6s$  &    $-$9      &    118\,(24)
                     &                             \\
$^1D_2$ &  $5d6s$  &     4        &    150
                     &                              \\
$^1S_0$ &  $6s7s$  &    2072      &
                     &                              \\
$^3S_1$ &  $6s7s$  &    2030      &    0.8
                     &                              \\
\hline $^3P_0^o$ & $6s6p$ &    252\,(25)   &
                     &                                  \\
$^3P_1^o$ & $6s6p$ &    278\,(15)    &    24.3\,(1.5)
                     &    24.06\,(1.37) \tablenotemark[2]    \\
                     &&&& 24.26\,(0.84)  \tablenotemark[3]     \\
                     &&&& 23.33\,(0.52)  \tablenotemark[4]     \\
$^3P_2^o$ & $6s6p$ &     383\,(32)   &    $-$76\,(6)
                     &                                   \\
$^1P_1^o$ & $6s6p$ &     501\,(200)    &    $-$118\,(60)
                     &  $-$57.4\,(5.6) \tablenotemark[2]    \\
\end{tabular}
\tablenotemark[1]{Ref.~\cite{Budk},}
\tablenotemark[2]{Ref.~\cite{Rinkleff},}
\tablenotemark[3]{Ref.~\cite{Kulina},}
\tablenotemark[4]{Ref.~\cite{Li}.}

\end{table}
\end{document}